\documentclass[preprint,12pt]{elsarticle}

\usepackage{microtype}
\usepackage{amsmath}
\usepackage{amssymb}
\usepackage[english]{babel}
\usepackage{latexsym}
\usepackage{mathrsfs}
\usepackage{graphicx}
\usepackage{dcolumn}
\usepackage{epsfig}
\usepackage{graphics}
\usepackage{color}
\usepackage{bm}
\usepackage{units,nicefrac}

\journal{Physica A}

\begin{document}

\begin{frontmatter}



\title{Is there a one-to-one correspondence between interparticle interactions and physical properties of liquid?}

\author[kpfu]{Anatolii V. Mokshin}
\ead{anatolii.mokshin@mail.ru}

\author[kpfu]{Roman A. Khabibullin}

\affiliation[kpfu]{organization={Department of Computational Physics, Institute of Physics, Kazan Federal University},
    city={Kazan},
    postcode={420008},
    country={Russia}}

\begin{abstract}
In this study, we present the original method for reconstructing the potential of interparticle interaction from statistically averaged structural data, namely, the radial distribution function of particles in many-particle system. This method belongs to a family of machine learning methods and is implemented through the differential evolution algorithm.
As demonstrated for the case of the Lennard-Jones liquid taken as an example, there is no one-to-one correspondence between structure and potential of interparticle interaction of a many-particle disordered system at a certain thermodynamic state.  Namely, a whole family of the Mie potentials determined by two parameters $p_{ 1 }$ and $p_{ 2 }$ related to each other according to a certain rule can reproduce properly a unique structure of the Lennard-Jones liquid  at a given thermodynamic state. It is noteworthy that this family of the potentials quite correctly reproduces for the Lennard-Jones liquid the transport properties (in particular, the self-diffusion coefficient) over a temperature range as well as the dynamic structure factor, which is one of the key characteristics of the collective dynamics of particles.
\end{abstract}



\end{frontmatter}


\section{Indroduction \label{sec: introduction}}

To reproduce most of the physical properties of a many-particle system, it is necessary to know its Hamiltonian, namely, the part of the Hamiltonian that defines the interparticle interaction. If the potential of interparticle interaction is known, then it is possible to calculate microscopic collective dynamics, to determine the transport coefficients (self-diffusion, viscosity, thermal conductivity), to compute the phase diagram as well as the characteristics of various phase transitions of the system under consideration~\cite{Landau_Stat_mech,Evance_Morrice_Stat_Mech}. For simple many-particle systems, the interaction potential may be initially known. Typical example of such systems is the system of interacting atoms/molecules of inert gases, where the interparticle interaction is reproduced by the Lennard-Jones potential~\cite{LJ,Kirkwood_1946}. Another example of simple system is a model single-component plasma, where the effective interparticle interaction is given by the Coulomb-type potential or the Yukawa potential~\cite{Hansen2006,Fairushin2020,AVM/IIF/IMT_PRE_2022,Likos_1}. However, for the overwhelming majority of real physical systems, it is necessary to solve the problem of finding the interaction energy $u(\mathbf{r})$ of particles as a function of radius vector $\mathbf{r}$ (or distance $r = |\mathbf{r}|$ for homogeneous isotropic case).

There are two theoretical concepts to recover the potential~$u(\mathbf{r})$ of a natural many-particle system. In accordance with the first and apparently most popular concept, the potential $u(\mathbf{r})$ is constructed in such a way that the energies of various configurations of the system and the resulting forces acting on each particle of the system coincide with the energies  and forces obtained from the \textit{ab initio} molecular dynamics simulations~\cite{Mueller2020,Behler_2016,Rowe2018,Deringer}. In this case, the search for the potential comes down either to finding the values of the parameters of a certain model potential, given in an analytical form~\cite{Ercolessi1994}, or to constructing the potential using neural networks~\cite{Blank1995,Ryltsev_2022}.

A feature of the second concept is that the search for the potential~$u(\mathbf{r})$ involves solving the so-called \textit{inverse problem}~\cite{Groetsch1999}. Here, it is required to restore the potential, assuming that values of some physical characteristics determined by this potential are known from experiment. These can be characteristics obtained from traditional experimental techniques for microscopy, neutron or X-ray diffraction, or from a numerical experiment such as molecular dynamics simulations or Monte-Carlo simulations.
The simplest implementation of the this approach assumes that the potential is constructed on the basis of known experimental data on the \emph{structure of the system}~\cite{McGreevy1988, Lyubartsev1995,Youngs_2019,Levesque_1,Chang_1}. The inverse problem in this case implies recovering the potential from the structural data and it is solved by the iteration (brute force) method as applied to some coarse-grained model of the potential~$u(\mathbf{r})$ with variable parameters. The corresponding enumeration is carried out until the required correspondence is reached in the values of a structural characteristic obtained with trial modification of the potential and measured experimentally~\cite{Soper_1,Chan2019,Thaler_2021}.
With the development of machine learning methods~\cite{Rodrigues2022,Jiang2019,Stoico2008,Grigorenko2000,Yu2020,Yadav2020, Sofos2022, Alam2022, Papastamatiou2022}, this approach is being improved in such a way that solving the problem requires fewer processing power as well as less computing time.  In this paper, we propose an original method for recovering the interparticle interaction potentials based on the differential evolution algorithm, which also belongs to the class of machine learning algorithms~\cite{Storn1997}. The efficiency of this method is demonstrated by the recovering the Lennard-Jones potential under the condition that the structural data~--- the radial distribution function $g(r)$ of the particles in this system~--- is known initially.

It is important to note that the following questions can be formulated in relation to any recovered potential of interparticle interaction. Is this recovered potential unambiguously determined, or are its various modifications possible, which will reproduce the experimental structure with acceptable accuracy? How correctly does this recovered potential reproduce other physical characteristics of the system under consideration (for example, the transport coefficients~--- self-diffusion, viscosity, thermal conductivity; the characteristics of collective particle dynamics~--- speed of sound propagation, sound attenuation coefficient, etc.)? Is it sufficient to rely only on structural data to recover the potential $u(\mathbf{r})$?  The purpose of this study is to provide answers to these questions.

The given study presents an original method for reconstruction of the interaction potential in a many-particle system. Although the main results are demonstrated on the example of a Lennard-Jones fluid, the method is general and can be applied to arbitrary systems of interacting particles in thermodynamic equilibrium (for example, equilibrium fluids of arbitrary physical nature). In particular, the method can be applied to estimate the interatomic interaction in heavy metal melts, where solving this problem on the basis of quantum mechanics calculations is laborious~\cite{Kruglov2019, Wang2022, Nghia2022, Beeler2021, Migdal2021}. In addition, it can be used to reconstruct effective coarse-grained potentials for the case of polymer melts and other disordered systems~\cite{Clark2013, Jin2022, Joshi2021}.

\section{Differential evolution algorithm for recovering the interparticle interaction potentials \label{sec: method}}

\subsection{Statement of the problem}

Let us assume that for an investigated many-particle homogeneous isotropic system, such as a liquid or an amorphous solid, structural data are initially known, and these structural data are specified by the radial distribution function $g(r)$ of particles. The system under study can be a real physical system or a model system, where the interaction of particles is specified by a certain potential~$u(r)$. Consequently, the structural data can be experimental or calculated based on the results of molecular dynamics simulations or Monte Carlo simulations with a given potential. The radial distribution function corresponding to these structural data is known. This function has the meaning of an \textit{input quantity} in the given problem and will be denoted as $g_{ \text{obj} }(r)$, where the subscript \underline{obj} means objective.

Further, our goal is to recover the potential of interparticle interaction $u(r)$. Let us assume that it belongs to the family of the potentials $\overline{u}_{ \mathbf{p} }(r)$, i.e.
\begin{equation} \nonumber
    u(r) \in \overline{u}_{ \mathbf{p} }(r),
\end{equation}
the analytical form of which are specified by the set of parameters
\begin{equation} \label{eq: parameters_p}
    \mathbf{p} = \{ p_{ 1 }, p_{ 2 }, \ldots, p_{ n } \},\quad n \in \mathcal{N}.
\end{equation}
The values of the parameters $\{ p_{ 1 }, p_{ 2 }, \ldots, p_{ n } \}$ are initially unknown. However, some general form of the potential $u_{ \mathbf{p} }(r)$ can be chosen taking into account the nature of the many-particle system. So, for example, it could be a system of point charges (electrons, ions), a system of dipoles or colloid particles, etc. Then, the parameters $\{ p_{ 1 }, p_{ 2 }, \ldots, p_{ n } \}$ can be the exponents in attractive or repulsive contributions, the weights of these contributions, the characteristics of selected directions in particle interactions in the case of nonspherical potentials, etc. Thus, the problem of recovering the potential is reduced to determining the values of these parameters at which the model potential $u_{ \mathbf{p} }(r)$ will correctly reproduce the structure of the system, namely, the `experimental' radial distribution function $g_{ \text{obj} }(r)$.

The exact relationship between the interaction potential $u(r)$ and the distribution function $g_{ \text{obj} }(r)$ is known only for the case of a rarefied gas~\cite{Fisher_book,Boon/Yip,Barker1976}:
\begin{equation} \label{eq: rarefied_gas}
    g_{ \text{obj} }(r) = \exp\left[ -\frac{u(r)}{k_{ \text{B} } T} \right].
\end{equation}
For the condensed phases such as a high-dense gas, liquid and amorphous solid, a correspondence between the quantities $g(r)$ and $u(r)$ is unknown, and it is not possible to calculate \emph{analytically} the interaction potential $u(r)$ on the basis of the known function $g_{ \text{obj} }(r)$. On the other hand, the radial distribution function $g(r)$ itself can be accurately calculated using configuration data resulted from molecular dynamics or Monte-Carlo simulations, if the potential $u(r)$ is specified for a given thermodynamic state of the system under study~\cite{Allen/Tildesley,Frenkel/Smit}. Thus, to restore the potential, we can use the following computational procedure. It is necessary to try different modifications of the potential $u_{ \mathbf{p} }(r)$ for molecular dynamics (or Monte-Carlo) simulations, determining the radial distribution function $g(r)$ and each time comparing the obtained trial function $g(r)$ with the objective function $g_{ \text{obj} }(r)$.
The procedure should be carried out until the required correspondence between functions $g(r)$ and  $g_{ \text{obj} }(r)$ is obtained.  Note that a modification of the potential is actually determined  by values of the parameters $\{ p_{ 1 }, p_{ 2 }, \ldots, p_{ n } \}$. Thus, in order to restore the potential in accordance with this procedure, we really need to sort out various combinations of the values of its parameters $\{ p_{ 1 }, p_{ 2 }, \ldots, p_{ n } \}$ and find the optimal one.

Let us choose the quantity $R(\mathbf{p})$ determined by the following relation
\begin{equation} \label{eq: residual}
    R(\mathbf{p}) = \frac{1}{\mathcal{N}} \sum_{ i = 1 }^{ \mathcal{N} } \left[ g_{ \text{obj} }(r_{ i }) - g_{ \mathbf{p} }(r_{ i }) \right]^{ 2 },
\end{equation}
as a measure of similarity (or difference) between the trial function $g(r)$ and the objective function $g_{ \text{obj} }(r)$. Here, $\mathcal{N}$ is the number of segments in $r$-axis, and $i$ is the label of a segment. In fact, the quantity $R(\mathbf{p})$ is the so-called mean squared deviation or weighted residual sum of squares known in statistics~\cite{Draper1999}, and it can only take non-negative values. If $R(\mathbf{p}) = 0$, then both the functions $g(r)$ and $g_{ \text{obj} }(r)$ coincide. The more these functions differ, the greater value the quantity $R(\mathbf{p})$ takes.

Further, let us introduce a phase space formed by the set of parameters $\mathbf{p}$. Hypothetically,  each model potential from the family $\overline{u}_{ \mathbf{p} }(r)$ can be associated with a value of the quantity $R(\mathbf{p})$. The set of resulting values of the quantity $R(\mathbf{p})$ will yield a landscape (surface) in this phase space. Then, the location of the global minimum of this landscape, where the quantity $R(\mathbf{p})$ tends to zero, should be directly associated with the values of the parameters $\{ p_{ 1 }, p_{ 2 }, \ldots, p_{ n } \}$ of the sought model potential $u(r)$. Finding this global minimum in the landscape is the same as finding  the potential $u(r)$, which is  capable to reproduce the experimental structure characteristic $g_{ \text{obj} }(r)$ with a required accuracy. Thus, we have the typical optimization problem that can be solved by means of the differential evolution algorithm~\cite{Storn1997}.

\subsection{Adapted differential evolution algorithm and its implementation}

The first step of the proposed computational scheme is to determine the family of model potentials, within which the search for the required model will be carried out. First of all, it is necessary to define some general model $\overline{u}_{ \mathbf{p} }(r)$, the free parameters $\mathbf{p}$ that set the model $\overline{u}_{ \mathbf{p} }(r)$, as well as the range of acceptable values of these parameters. An important point is the number of free parameters $\mathbf{p}$ and the range of their acceptable values. A relatively small number of parameters with narrow ranges of acceptable values simplifies the optimization process and increases the chances of achieving convergence of the values of free parameters with limited computing resources. On the other hand, a larger number of parameters significantly increases  a flexibility of parameterization and increases the chances of finding a solution.

The differential evolution algorithm as applied to the problem of potential recovering includes two computing procedures.

The first procedure is associated with finding the equilibrium configurations of many-particle system in a given thermodynamic state. These configurations should be generated using the test model potential of interparticle interaction. They will be required for the subsequent calculation of the radial distribution function $g(r)$ and comparison with the objective function $g_{ \text{obj} }(r)$. Consequently, it is convenient here to use a computational program or package that supports molecular dynamics or Monte-Carlo simulations. For example, this could be the LAMMPS package~\cite{Plimpton1995}, which implements molecular dynamics simulations and is distinguished by high computational speed. It supports for calculations on GPU, simulations in various statistical ensembles ($NVE$, $NVT$, $NpT$), as well as on-the-fly calculations of various structural and transport properties.

\begin{figure}
    \centering
    \includegraphics[keepaspectratio,width=0.3\linewidth]{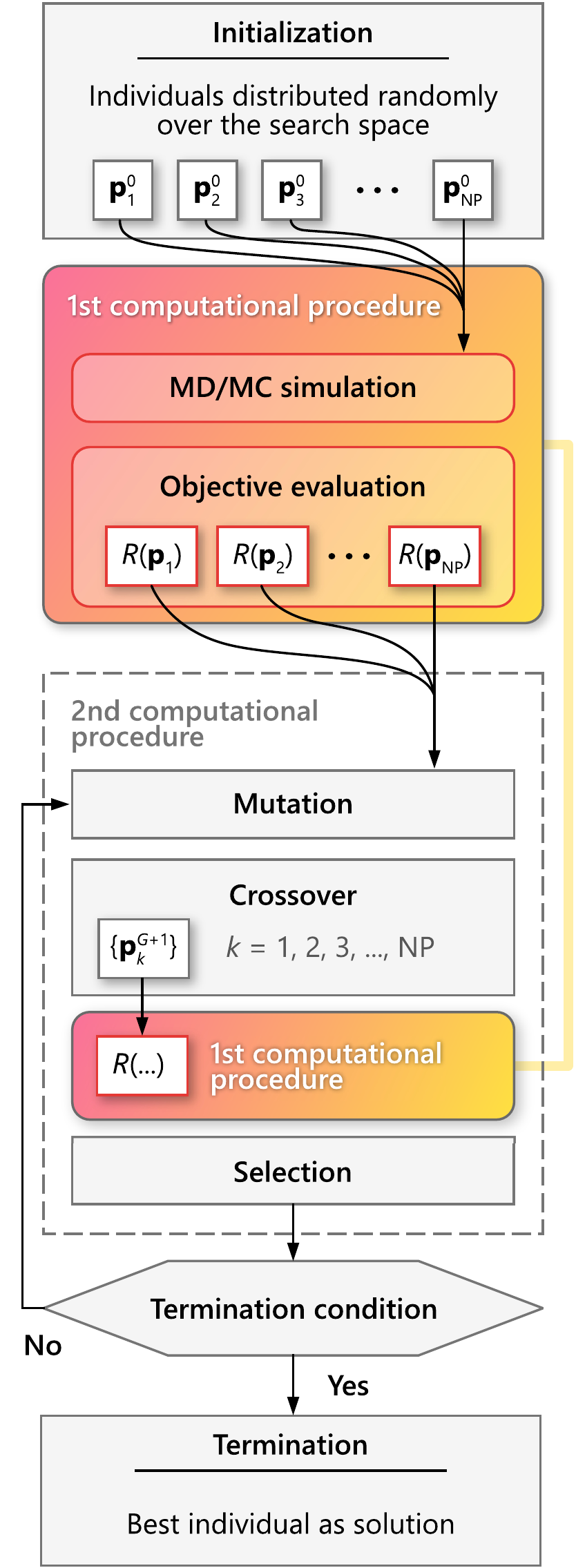}
    \caption{Flowchart of the differential evolution method.}
    \label{fig: flowchart}
\end{figure}
The second procedure involves computations in the framework of the adapted differential evolution algorithm, the main elements of which are given in Fig.~\ref{fig: flowchart}. In accordance with the standard terminology of evolutionary algorithms, one can say that this computational procedure involves the following: formation of generations of individuals, as well as operations of mutation, crossover and selection of individuals~\cite{EA_base}. So, let us define these basic terms and operations of the differential evolutionary algorithm methodology.

\begin{figure*}
    \centering
    \includegraphics[keepaspectratio,width=0.8\linewidth]{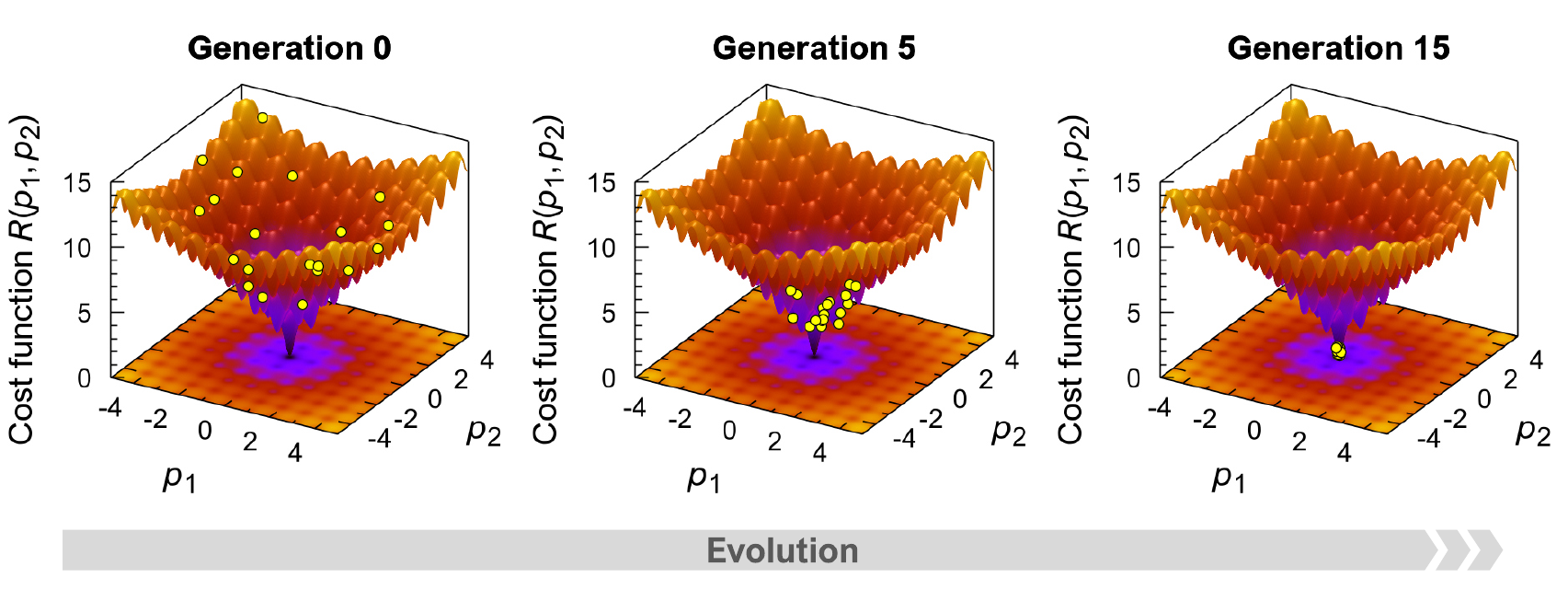}
    \caption{Finding the global minimum of the landscape $R(p_{ 1 }, p_{ 2 })$ given by the Ackley function with the two parameters $p_{ 1 }$ and $p_{ 2 }$. Initially, in the population of the zeroth generation, the individuals shown in yellow markers ($\bullet$) are evenly and randomly scattered across the landscape. Despite the complex shape of the landscape $R(p_{ 1 }, p_{ 2 })$, containing many local minima, the adapted evolutionary algorithm successfully brings all individuals to the global minimum at the point $R(0, 0)$. An individual corresponding to the desired solution is found in the population of the $15$th generation.}
    \label{fig: Ackley}
\end{figure*}
By \textit{individual}, we mean a model potential $u_{ \mathbf{p}_{ i } }(r)$ from the initially given family of potentials $\overline{u}_{ \mathbf{p} }(r)$. Here, $i$ is the label of an individual with a unique \textit{genome} $\mathbf{p}_{ i }$, which is a set of the parameters with specific numerical values, i.e.
\begin{equation*}
    \mathbf{p}_{ i } = \{ p_{ 1 }^{ (i) }, p_{ 2 }^{ (i) }, \ldots, p_{ n }^{ (i) } \}.
\end{equation*}
Thus, $p_{ n }^{ (i) }$ has the meaning of a \textit{gene}. From mathematical point of view, the genome $\mathbf{p}_{ i }$ can be viewed as a vector with components $p_{ 1 }^{ (i) }$, $p_{ 2 }^{ (i) }$, $\ldots$, $p_{ n }^{ (i) }$. Therefore, all the standard mathematical operations with vectors are also applicable to it. Then, the quantity $n$ has the meaning of the dimensionality of the search space.
A set of individuals form a \textit{population} of a certain \textit{generation} $G$; and the population size is denoted as $NP$. The size of the population is determined by the dimensionality of the search space $n$~\cite{Storn1997}. As was found earlier (see, for example, work~\cite{Storn1997}, p. 356), there is a rule of thumb which establishes a quantitative correspondence between the dimensionality of the search space $n$ and the minimum number of individuals in the population $NP$:
\begin{equation}
    NP = (5 \div 10)\, n\,.
\end{equation}
The minimal permissible value of $NP$ is 4. We note that the term generation is, in a sense, a characteristic of a population: thus, there may be a population of a parent or a child generation. Therefore, in some cases it is convenient and appropriate to use the term generation instead of population. Individuals of each generation undergo operations of mutation, crossover and selection.

\textit{Mutation.} -- Mutation is defined as the mathematical operation applied to three random genomes $\mathbf{p}_{ k }^{ G }$, $\mathbf{p}_{ l }^{ G }$ and $\mathbf{p}_{ m }^{ G }$ of the individuals of the same population $G$ in accordance with the following vector addition rule:
\begin{equation} \label{eq: mutation}
    \mathbf{p}_{ k }^{ \widetilde{G} + 1 } = \mathbf{p}_{ k }^{ G } + F \left(
        \mathbf{p}_{ l }^{ G } - \mathbf{p}_{ m }^{ G } \right).
\end{equation}
Here, the weight coefficient $F$ known also as the \textit{mutation factor} can take some a chosen non-integer value from the range $[0, 2]$ as indicated in Ref.~\cite{Price2005}. The higher value of $F$, the more perturbations/changes are introduced into the genome. As found in Ref.~\cite{Storn1997}, values of $F$ smaller than 0.4 or larger than 1 are effective only in specific cases. This operation is applied to the genomes of all the individuals of the population $G$. The operation is considered acceptable if the obtained values of the genome parameters $\mathbf{p}_k$ of the individual $u_{ \mathbf{p}_{ k } }(r)$ of the new mutated generation $(\widetilde{G} + 1)$ do not go beyond the range of permissible values.
Such an implementation of mutation was proposed in the original work on differential evolution and it is essentially a reproduction of a similar operation in the Nelder-Mead method~\cite{Price2005} (see, for example, Eq. (1.25) in~\cite{Price2005}). It should be noted that to improve population diversity, the mutation operation given by equation (\ref{eq: mutation}) can be modified so as to be performed at a time on a larger number of individuals, if the population size $NP$ is large enough~\cite{Price1996}.

\textit{Crossover.} -- Crossover is the operation carried out with the genomes of a pair of individuals, one each from the parental $G$ and the new mutated $(\widetilde{G} + 1)$ generations. Let us show this operation by the example of the formation of the $i$-th individual for the new generation $(G + 1)$:
\begin{equation}
    p_{ n }^{ G + 1 } =
    \begin{cases}
        p_{ n }^{ \widetilde{G} + 1 } & \text{if $\xi \leqslant C_{ R }$ or $n = n_{ \text{rand} }$\,,} \\
        p_{ n }^{ G } & \text{otherwise.}
    \end{cases} \label{eq: crossover}
\end{equation}
Here, $\xi$ is a random non-integer number from $[0, 1)$. All the individuals of both the generations $G$ and $(\widetilde{G} + 1)$ must undergo this operation. Eq.~(\ref{eq: crossover}) sets that all the components of a vector $\mathbf{p}_{ k }^{ G }$ of the parental generation $G$ are replaced with a certain probability $C_{ R }$ by the components of the new mutated vector $\mathbf{p}_k^{ \widetilde{G} + 1 }$, while a one component, chosen at random and with the label $n_{ \text{rand} }$, is necessarily replaced by the corresponding component of the mutated vector. The quantity $C_{ R }$ is referred to as the \textit{crossover rate} and can take a non-integer value from the range $[0, 1]$. Sequential application of mutation and crossover operations ensures that all the individuals of the new generation $(G + 1)$ will be distinct from the individuals of the parent generation $G$. The greater the value of $C_{ R }$, the more substantial is the change in the individuals. Solving the problem of restoring potential is computationally challenging, so rapid evolution of generations is necessary. Therefore, it is reasonable to set large values of $C_{ R }$, close to 1.

\textit{Selection.}~--- The last stage, as a result of which the child generation is finally formed, is associated with the \textit{selection} operation. The selection operation must ensure the evolution of generations in a required direction. Consequently, those individuals should be selected whose characteristics are closest to the required ones. In the problem under consideration, selection is carried out on the basis of a parameter $R(\mathbf{p}_{ j })$ determined for each $j$th individual [see Eq.~(\ref{eq: residual})]. The evolution should lead to the individuals with the lowest values of the parameter $R$, which, in the methodology of evolutionary algorithms, has a meaning of a \textit{fitness function} or \textit{cost function}~\cite{Price2005}. To perform the selection operation, it is necessary to consider once again the pairs of individuals, one for the parental $G$-generation and for the $(G + 1)$-generation obtained as a result of mutations and crossovers. For the child generation, an individual is selected from these two, for which the cost function $R$ takes the smallest value.

Flowchart of the adapted differential evolution algorithm is given in Fig.~\ref{fig: flowchart}. The initial population, i.e. the zero generation population, is created by selecting random individuals in the phase space (search space). Note that the probabilities of choosing one or another individual are the same. This is followed by the first computational procedure associated with finding equilibrium configurations for individuals in a given generation. Here, for all the individuals, the cost function $R$ is calculated. After that, the condition for termination of the computational algorithm is checked. If the condition is not met, then there is a transition to the second computing procedure with the simulation of the evolution process and a subsequent return to the first computing procedure. The algorithm is performed until an individual with the required properties is found.

In fact, this algorithm should implement the search for the global minimum of the landscape $R(\mathbf{p})$. The more complex the geometry of the landscape $R(\mathbf{p})$, the more difficult it is to carry out such a search. Fig.~\ref{fig: Ackley} shows the results of testing the adapted differential evolution algorithm as applied to finding the global minimum of the model surface given by the Ackley function~\cite{Ackley1987}. Notice that the Ackley function is a non-convex function containing many local minima, and it is usually used to test performance for optimization algorithms. In our test case, we define the cost function $R(\mathbf{p})$ as the Ackley function of two variables $p_{ 1 }$ and $p_{ 2 }$:
\begin{eqnarray} \label{eq: Ackley}
    R(p_{ 1 }, p_{ 2 }) & = & -20 \exp\left[ -\frac{1}{5} \sqrt{\frac{\left( p_{ 1 }^{ 2 } + p_{ 2 }^{ 2 } \right)}{2}} \right] \\
    & & - \exp\left[ \frac{\cos{2\pi} p_{ 1 } + \cos{2\pi} p_{ 2 }}{2} \right] + e + 20\,, \nonumber
\end{eqnarray}
which has the global minimum
\begin{equation*}
    R(0,\, 0) = 0\,.
\end{equation*}
In Eq.~(\ref{eq: Ackley}), $e$ is the Euler's number, and all the numerical coefficients are the same as in the original work~\cite{Ackley1987}.
As seen in Fig.~\ref{fig: Ackley}, the fifteenth generation population converges at the point $R(0,\, 0)=0$, indicating that the adapted differential evolution algorithm successfully finds the global minimum. Note that the algorithm was performed for a population of the size $NP = 20$. Following the results of Ref.~\cite{Centeno}, the mutation factor $F$ and crossover rate $C_{ R }$ were taken to be $0.44$ and $0.88$, respectively.

\section{Recovering the Lennard-Jones potential \label{sec: LJ}}

\subsection{Statement of the problem \label{sec: Problem_statement}}

Let us check the efficiency of the above method for the problem of restoring the Lennard-Jones potential
\begin{equation}
    u_{ \text{LJ} }(r) = 4\epsilon \left[ \left( \frac{\sigma}{r} \right)^{ p_{ 1 } } - \left( \frac{\sigma}{r} \right)^{ p_{ 2 } } \right]
\label{eq: LJ}
\end{equation}
\begin{equation*}
    \mathrm{with}\quad p_{ 1 } = 12\quad \mathrm{and}\quad p_{ 2 } = 6
\end{equation*}
from the known radial distribution function $g(r)$ for some thermodynamic ($\rho,\, T$)-state.  Here, $\epsilon$ and $\sigma$ are the characteristic scales of energy and length given by the potential; and, throughout the section, all physical quantities will be defined in terms of these characteristics.

The Lennard-Jones potential is a special case of the Mie potential~\cite{Mie1903}, which is a linear combination of repulsive and attractive contributions with arbitrary power dependencies:
\begin{equation}
    \overline{u}_{ p_{ 1 }, p_{ 2 } }(r) = \frac{p_{ 1 }}{p_{ 1 } - p_{ 2 }} \left( \frac{p_{ 1 }}{p_{ 2 }} \right)^{ \frac{p_{ 2 }}{p_{ 1 } - p_{ 2 }} } \epsilon \left[ \left( \frac{\sigma}{r} \right)^{ p_{ 1 } } - \left( \frac{\sigma}{r} \right)^{ p_{ 2 } } \right]
\label{eq: Mie}
\end{equation}
\begin{equation*}
    \mathrm{with}\quad p_{ 1 }, p_{ 2 } > 0,
\end{equation*}
and
\begin{equation*}
    u_{ \text{LJ} }(r) \in \overline{u}_{ p_{ 1 }, p_{ 2 } }(r).
\end{equation*}
The concrete analytic form of the Mie potential is given by the pair of parameters $p_{ 1 }$ and $p_{ 2 }$.
Thus, if the function $g(r)$ for the Lennard-Jones system is known, then the problem of restoring the Lennard-Jones potential naturally reduces to finding the numerical values of the parameters $p_{ 1 }$ and $p_{ 2 }$ of the Mie potential or to finding the point $R(p_{ 1 }, p_{ 2 })$ corresponding to a global minimum on the landscape $\mathcal{R}^{ 2 }$ formed by the parameters $p_{ 1 }$ and $p_{ 2 }$.

\subsection{Relevant information about the system, calculations and simulation details}

\begin{figure}[h!]
    \centering
    \includegraphics[keepaspectratio,width=0.5\linewidth]{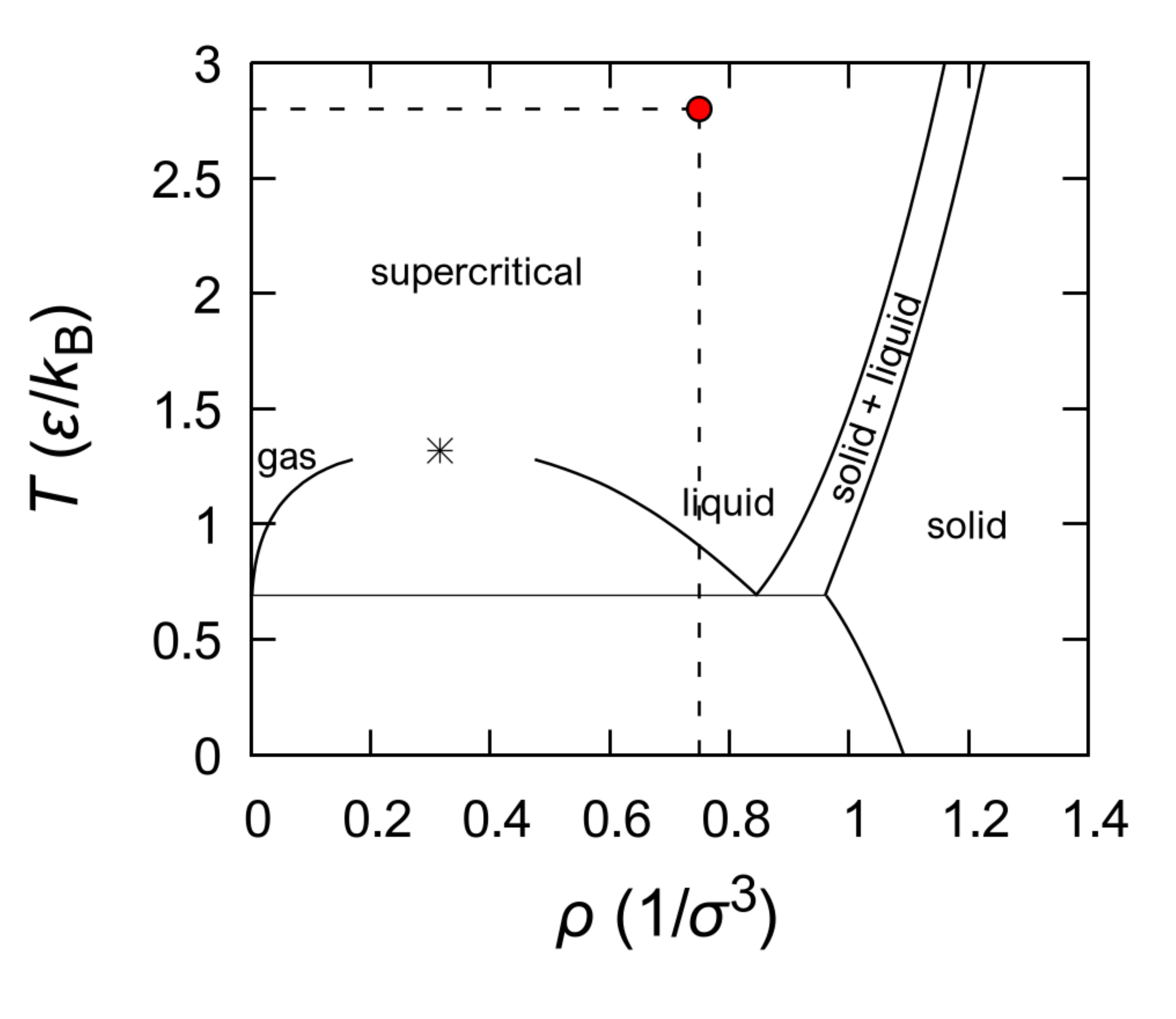}
    \caption{Phase diagram of the Lennard-Jones system in variables $\rho$ and $T$ reproduced on the basis of data from Refs.~\cite{Stephan_2019} and \cite{Schultz_JCP}. Red filled circle ($\bullet$) corresponds to the system's state considered in this study.}
    \label{fig: phase_diagr_LJ}
\end{figure}

In the case of simple monatomic systems, such as the Lennard-Jones and/or Mie system, the thermodynamic states with the most complex homogeneous structure arise for the equilibrium liquid and supercritical fluid phases~\cite{Trachenko_RPP}. For this reason, in this study, we consider the state with the temperature $T = 2.8\, \epsilon/k_B$ and the density $\rho = 0.75\, \sigma^{ -3 }$, that corresponds for the case of the Lennard-Jones system to supercritical fluid at the pressure $p = 6.6\, \epsilon/\sigma^{ 3 }$. The ($\rho,\, T$) phase diagram of the Lennard-Jones system is given in Fig.~\ref{fig: phase_diagr_LJ}. Note that for this isobar $p = 6.6\, \epsilon/\sigma^{ 3 }$ the melting temperature (liquidus) of the Lennard-Jones system is estimated to be $T_m \simeq 1.1\, \epsilon/k_B$, whereas the isotherm $T = 2.8\, \epsilon/k_B$ intersects the melting line in the ($p,\, T$) phase diagram at the pressure $p_m \simeq 6.0\, \epsilon/\sigma^{ 3 }$. Obviously, when the values of the parameters $p_{ 1 }$ and $p_{ 2 }$ of the Mie potential change from the values corresponding to the Lennard-Jones potential, the melting line in the phase diagram will also change.
\begin{figure*}
    \centering
    \includegraphics[keepaspectratio,width=0.9\linewidth]{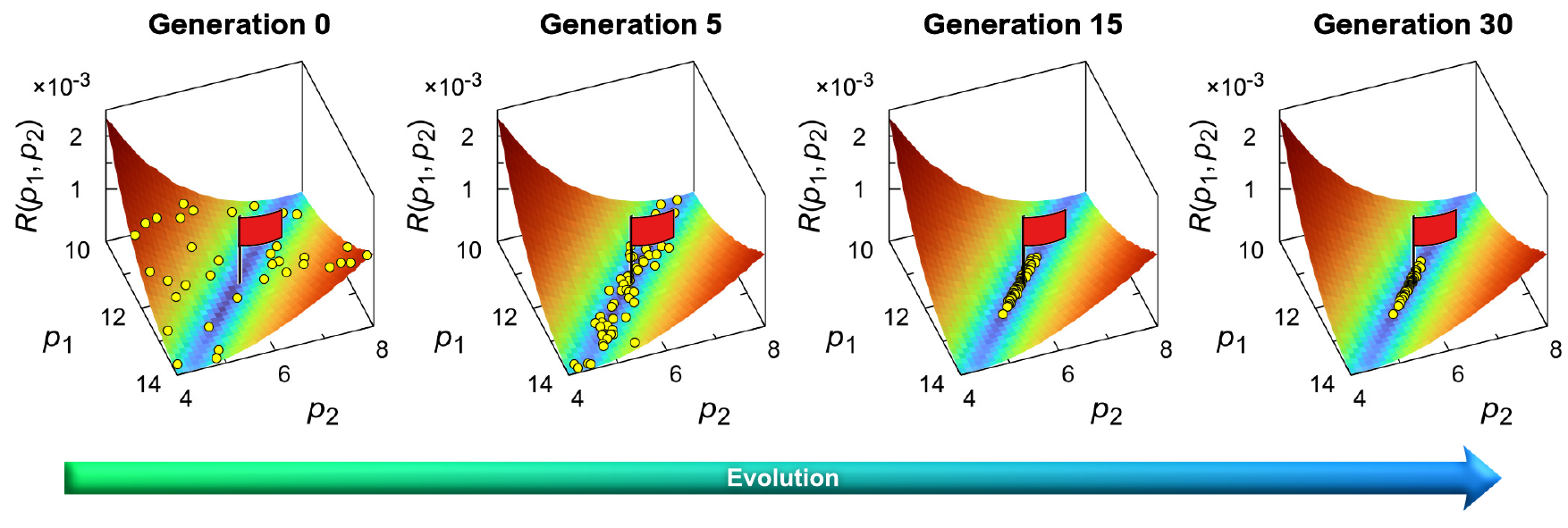}
    \caption{Finding the values of the parameters $p_{ 1 }$ and $p_{ 2 }$ of the Mie potential at which the Mie potential is able to reproduce the radial distribution function of a Lennard-Jones liquid at the temperature $T = 2.8\, \epsilon/k_B$ and the density $\rho = 0.75\, \sigma^{ -3 }$. The panels in the figure correspond to different generations of individuals—namely, zero, fifth, fifteenth, and thirtieth generations—after the start of evolution. Here, yellow circles indicate individuals, and the flag marks an individual with the gene $\{ p_{ 1 } = 12.0,\, p_{ 2 } = 6.0 \}$ corresponding to the desired solution. For clarity, the landscape $R(p_{ 1 }, p_{ 2 })$ is shown, restored using additional calculations, in which the entire ranges of both the parameters $p_{ 1 }$ and $p_{ 2 }$ were uniformly covered.
    }
    \label{fig: LJ_from_Mie}
\end{figure*}

For the chosen thermodynamic state of the Lennard-Jones fluid, the objective radial distribution function $g_{ \text{obj} }(r)$ was \textit{preliminarily} determined, which was subsequently used as an input quantity in the calculation procedures. The computational procedure related to the determination of the radial distribution function $g_{ p_{ 1 }, p_{ 2 } }(r)$ with the trial potential $u_{ p_{ 1 }, p_{ 2 } }(r)$ was performed in the framework of equilibrium molecular dynamics simulations in the $NpT$ ensemble. This ensemble is chosen for two reasons. First, such an ensemble makes it possible to realize conditions close to experimental ones, where temperature and pressure can be controlled. Second, molecular dynamics simulations in $NVE$ or $NVT$ ensembles with trial potentials can produce structurally heterogeneous systems (e.g., with pores and voids). In other words, no guarantee that any trial potential at $V, E = \mathrm{const}$ (or $V, T = \mathrm{const}$) conditions will produce the equation of state of the target system (namely, the equilibrium density of the target system). All simulations were performed for a cubic simulation box with $N =30\,000$ particles; and periodic boundary conditions were applied in all directions. The pressure was controlled by the Berendsen barostat~\cite{Allen/Tildesley}. To optimize the molecular dynamics simulations, we apply to the trial potential $u_{ p_{ 1 }, p_{ 2 } }(r)$ and the corresponding force $f_{ p_{ 1 }, p_{ 2 } }(r)$ the so-called shifted force technique given by the conditions~\cite{Toxvaerd2011}
\begin{subequations}
\begin{equation}
    u^{ \text{SF} }_{ p_{ 1 }, p_{ 2 } }(r) =
    \begin{cases}
        u_{ p_{ 1 }, p_{ 2 } }(r) - \left. (r - r_{ c }) \displaystyle{\frac{\partial}{\partial r}} u_{ p_{ 1 }, p_{ 2 } }(r) \right|_{ r = r_{ c } } \\ \\
        \hskip 1.5cm - u_{ p_{ 1 }, p_{ 2 } }(r_ { c }) & \hskip -1.8cm \text{if } r \leq r_{ c }; \\ \\
        0 & \hskip -1.8cm \text{if } r > r_{ c }
    \end{cases}
\end{equation}
and
\begin{equation}
    f^{ \text{SF} }_{ p_{ 1 }, p_{ 2 } }(r) =
    \begin{cases}
        f_{ p_{ 1 }, p_{ 2 } }(r) - f_{ p_{ 1 }, p_{ 2 } }(r_{ c }) & \text{if } r \leq r_{ c }; \\
        0 & \text{if } r > r_{ c }
    \end{cases}
\end{equation}
\end{subequations}
with the cutoff distance chosen as $r_{ c } = 5.0\, \sigma$.

The search for the values of the Mie potential parameters that would be able to reproduce the radial distribution function of the Lennard-Jones liquid was carried out in the following ranges of parameter values:
\begin{equation} \label{eq: ranges}
    p_{ 1 } \in [10.0, 15.0]\quad \text{and}\quad p_{ 2 } \in [3.0, 8.0],
\end{equation}
where the parameters $p_{ 1 }$ and $p_{ 2 }$ are real numbers specified with an accuracy of five decimal place.

The first (main) maximum of the objective radial distribution function $g_{ \text{obj} }(r)$ of the Lennard-Jones liquid is characterized by a shape close to Gaussian, which is typical for a system of particles interacting through the softcore potential. The exponent $p_{ 1 } = 15.0$ for the repulsion contribution taken as the largest in our considerations seems reasonable, since for larger values of $p_{ 1 }$ this contribution will approach the hard-sphere potential. Further, the selected maximum value $p_{ 2 } = 8.0$ of the attractive contribution corresponds to the dipole-quadrupole interaction, which strongly depends on distance $r$. Nevertheless, it is necessary to note that the ranges for values of both the parameters $p_{ 1 }$ and $p_{ 2 }$ can be taken even larger than in (\ref{eq: ranges}).

Thus, the search for the potential is carried out on the basis of the known radial distribution function~$g_{ \text{obj} }(r)$. It is assumed that the temperature and density corresponding to this function are known. Trial potentials of interparticle interaction are generated on the basis of the Mie potential given by expression~(\ref{eq: Mie}), where the parameters $p_{ 1 }$ and $p_{ 2 }$ can take values from the ranges defined by (\ref{eq: ranges}). Thus, the ranges given by (\ref{eq: ranges}) determine the specific region of the $\mathcal{R}^{ 2 }$ space within which the point $R(p_{ 1 }, p_{ 2 })$ will be searched.

Finally, it should be noted that the size of each population was sixty-four individuals (trial potentials), i.e. $NP = 64$, that is optimal from point of view of the available computing resources. As in the case of the Ackley function considered above, the mutation rate $F$ and the crossover rate $C_{ R }$ were taken equal to $0.44$ and $0.88$, respectively. These values of both the parameters are taken from~\cite{Centeno}; these values provide the fastest evolution of the population to the one containing the desired individual (solution). The choice of other acceptable values of these control parameters should not affect the final result.

\subsection{Results \label{sec: results}}

The applied algorithm finds succefully the desired individual corresponding to the Lennard-Jones potential. As seen from the scheme given in Fig.~\ref{fig: LJ_from_Mie}, the global minimum of the landscape~$R(p_{ 1 }, p_{ 2 })$ corresponding to the Lennard-Jones potential is achieved in the process of evolution with fifteen generations.  Further, in the process of evolution over ninety generations, the accuracy of the found parameters $p_{ 1 } = 12.0$ and $p_{ 2 } = 6.0$ is $\pm 10^{ -5 }$. All the individuals of the ``final generation'' are within a clear recognizable narrow range containing the desired individual $u_{ \text{LJ} }(r)$. This is evidence of the efficiency of the proposed algorithm and indicates the successful solution of the problem formulated above in Sec.~\ref{sec: Problem_statement}.

The results obtained lead to the question: Is there a one-to-one correspondence between the potential of interaction and the physical properties of the system? The answer to this question seems to be important even in the case of a simple monatomic system, where the interaction of an arbitrary pair of particles is determined only by their mutual distance, as in the case of the Lennard-Jones system. Let us try to get an answer to this question using the Lennard-Jones system as an example.

\begin{figure}[t]
    \centering
    \includegraphics[keepaspectratio,width=0.5\linewidth]{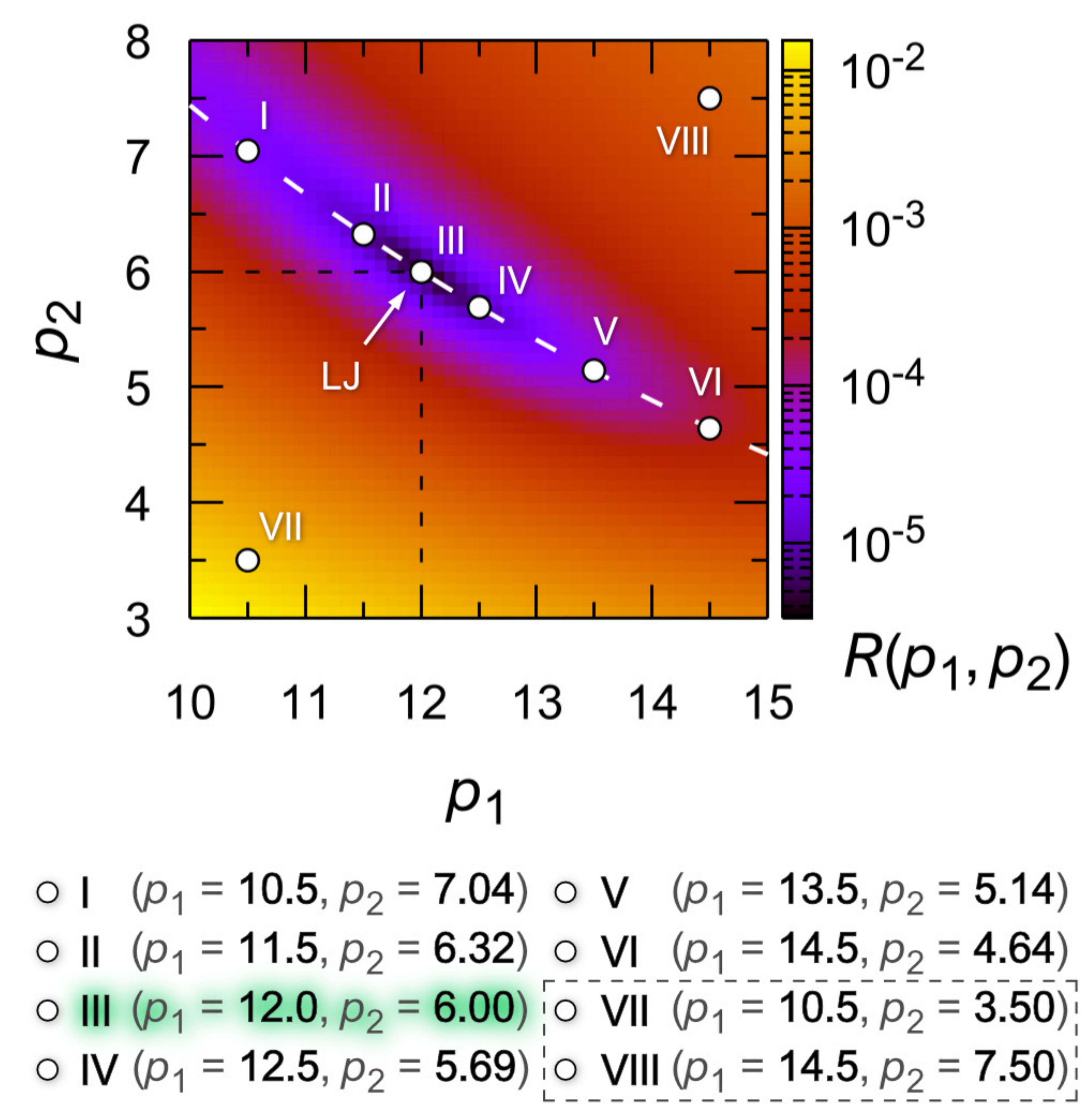}
    \caption{ Residual $R(p_{ 1 }, p_{ 2 })$ restored for the entire range of considered values of parameters $p_{ 1 }$ and $p_{ 2 }$. Circles with Roman numerals indicate combinations of Mie potential parameter values that define systems whose structural, dynamic, and transport properties will be further compared with the properties of the Lennard-Jones fluid. Here, the first six circles (or the points I-VI) correspond to the valley of the landscape, where the function~$R(p_{ 1 }, p_{ 2 })$ takes its minimum values. The circle with the number III corresponds to
    the Lennard-Jones system. The dashed line is the result of Eq.~(\ref{eq: correspondence}).
    }
    \label{fig: landscape_R}
\end{figure}
We start with the found residual $R(p_{ 1 }, p_{ 2 })$, which quantitatively characterizes the degree of correspondence between the structure of the Mie system and the structure of the Lennard-Jones fluid. The best correspondence will be for such values of the parameters $p_{ 1 }$ and $p_{ 2 }$ for which the landscape $R(p_{ 1 }, p_{ 2 })$ has a minimum. As seen in Fig.~\ref{fig: landscape_R}, there is a depression (minimum) on the smooth landscape~$R(p_{ 1 }, p_{ 2 })$, which is not localized, but is represented by a narrow valley. It is quite natural that in this valley there is the point III corresponding to the Lennard-Jones system. The orientation of the valley in the phase space $(p_{ 1 }, p_{ 2 })$ is such that, when moving along it, an increase in the parameter $p_{ 1 }$ is compensated by a decrease in the parameter $p_{ 2 }$, and \textit{vice versa}. As it turns out, such a correspondence between the parameters $p_{ 1 }$ and $p_{ 2 }$ results in a relatively conserved character of interparticle interaction, which is of the Lennard-Jones type. 
Let us choose six points (I -- VI) along the valley and two points (VII and VIII) maximally remote from the valley as shown in Fig.~\ref{fig: landscape_R} and consider further the structural, dynamic, and transport properties of the systems, which are determined by the values of the parameters of the Mie potential corresponding to these $(p_{ 1 }, p_{ 2 })$ points.

\begin{figure}[t!]
    \centering
    \includegraphics[keepaspectratio,width=0.5\linewidth]{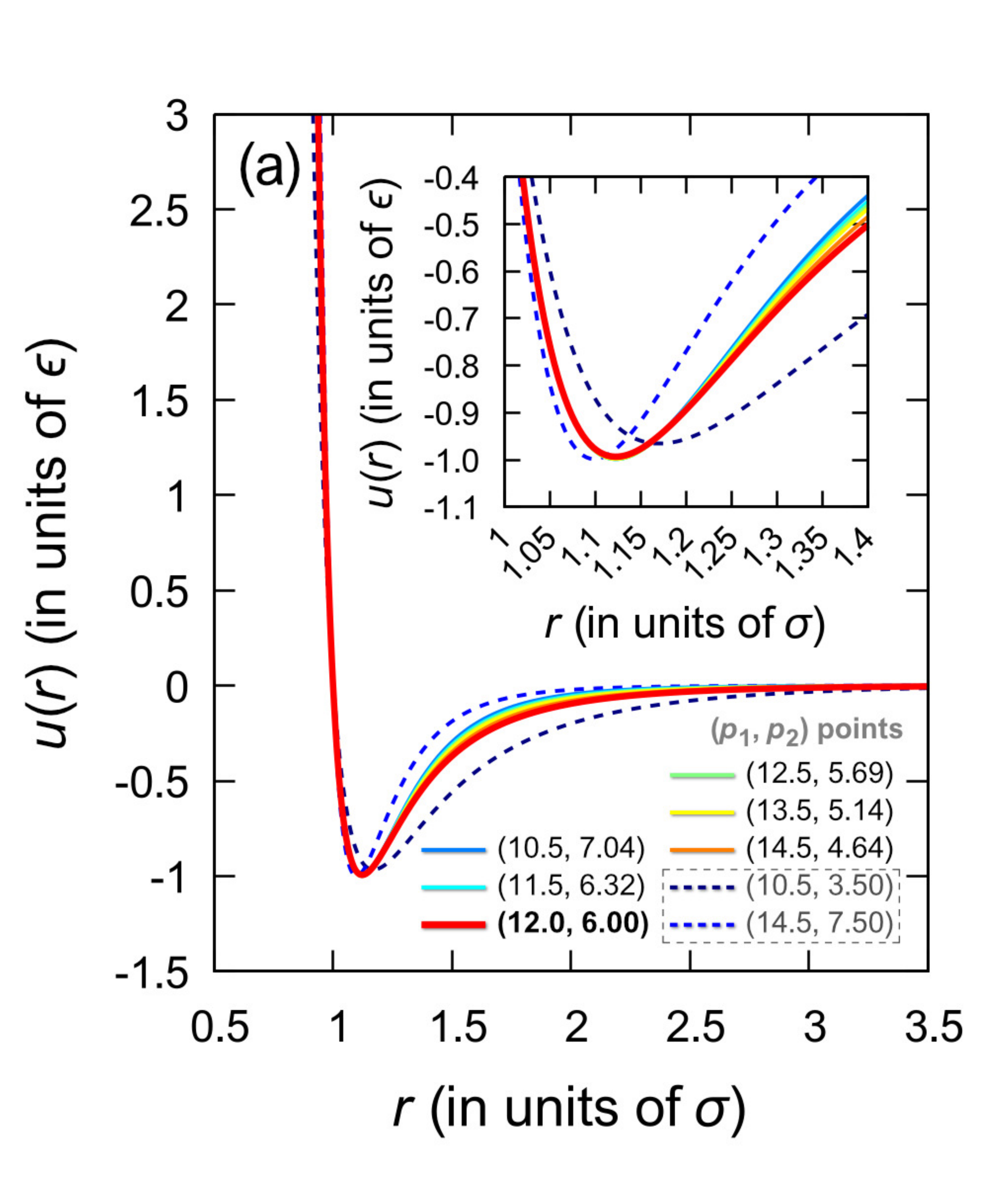}
    \includegraphics[keepaspectratio,width=0.5\linewidth]{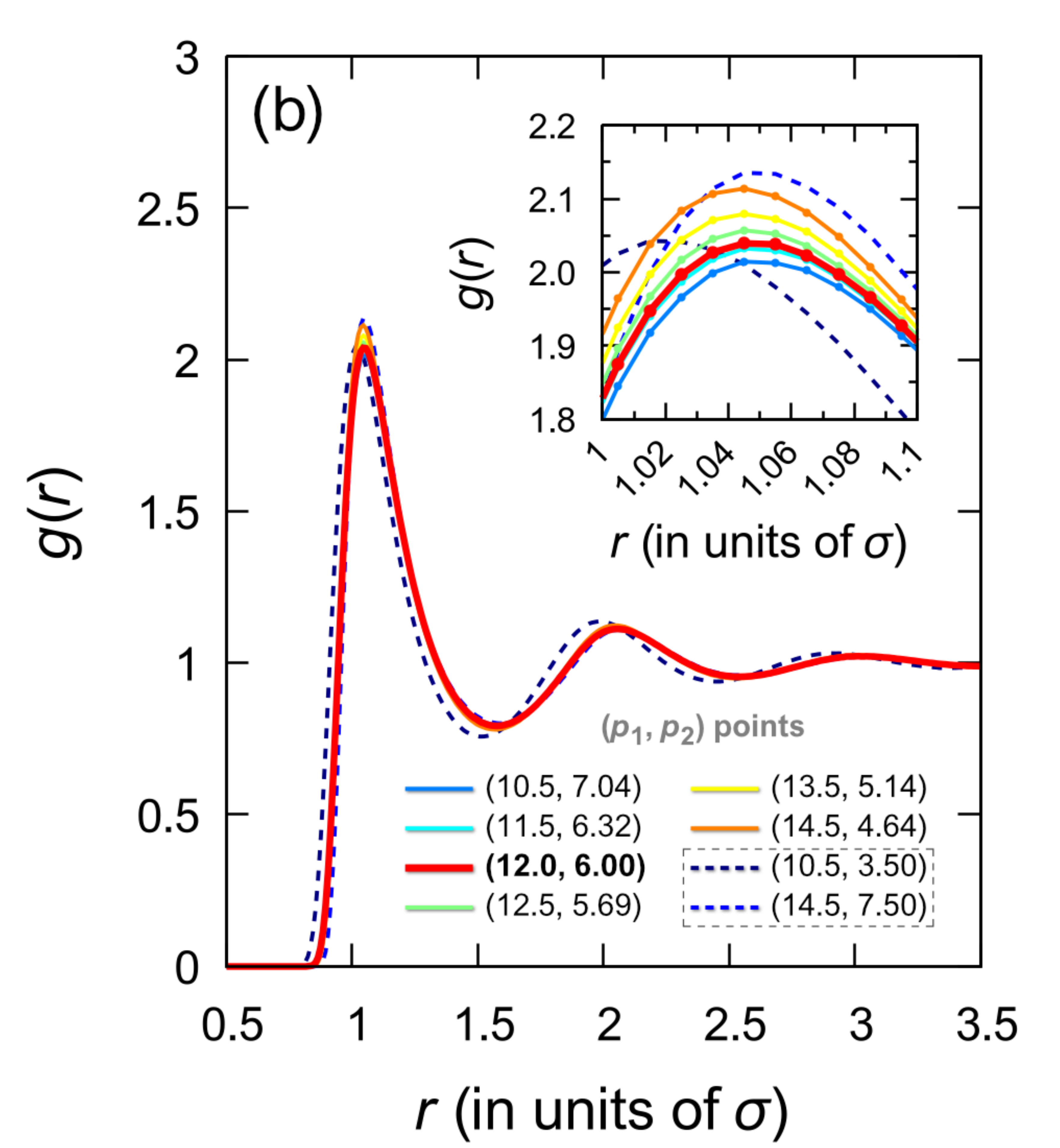}
\caption{(a) Mie potential for various eight$(p_{ 1 }, p_{ 2 })$ points. (b) Radial distribution function $g(r)$ computed by means molecular dynamics simulations with the Mie potentials for the same eight $(p_{ 1 }, p_{ 2 })$ points. All the functions $g(r)$ correspond to the same state with the temperature $T = 2.8\, \epsilon/k_{ \text{B} }$ and the density $\rho = 0.75\, \sigma^{ -3 }$.
    }
    \label{fig: structure_potential}
\end{figure}
In Fig.~\ref{fig: structure_potential}(a), the Mie potential is shown for the eight combinations of the parameters $p_{ 1 }$ and $p_{ 2 }$ defined in Fig.~\ref{fig: landscape_R}. For points $(p_{ 1 }, p_{ 2 })$, marked with symbols I-VI, the general form of the Mie potential is practically the same, and only at the distances $r \sim 1.6\, \sigma$, correlated with the outer boundary of the first coordination sphere, very slight differences are found. Further, the minimum in the Mie potential will be located as in the Lennard-Jones potential at the distance $r_{ 0 } = \sqrt[6]{2}\, \sigma$, if the following equality is satisfied:
\begin{equation}
    \left ( \frac{p_{ 1 }}{p_{ 2 }} \right )^{ 1 / (p_{ 1 } - p_{ 2 }) } = 2^{ 1 / 6 }.
\label{eq: LJ_MIe_eq}
\end{equation}
Solving this equation, we find the correspondence between the parameters of the Mie potential:
\begin{equation}
    p_{ 2 } = -\frac{6}{\ln{2}}\; W \left( -\frac{\ln{2}}{6} \frac{p_{ 1 }}{2^{\, p_{ 1 } / 6} } \right),
\label{eq: correspondence}
\end{equation}
\begin{figure}[t!]
    \centering
    \includegraphics[keepaspectratio,width=0.5\linewidth]{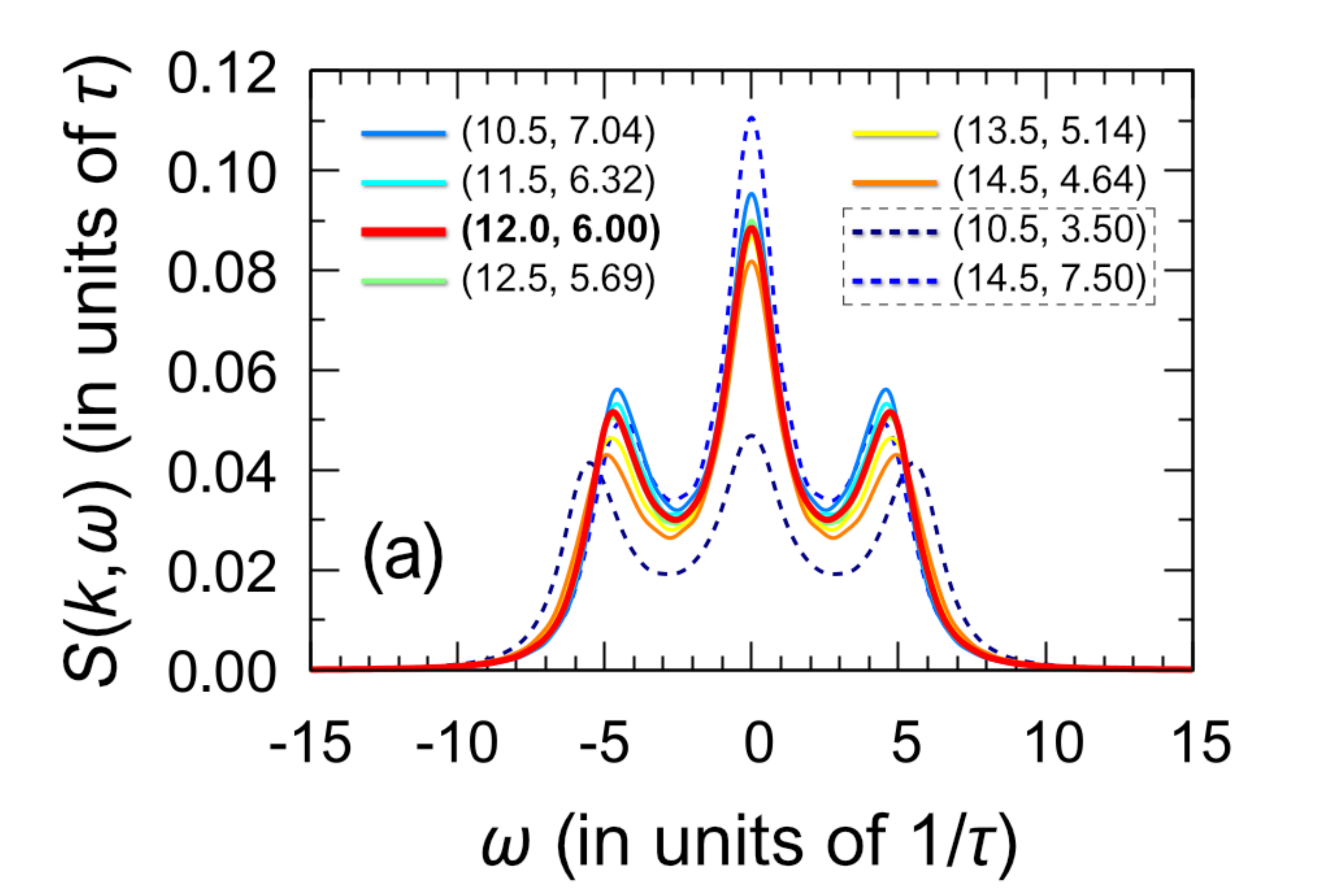}\\
    \includegraphics[keepaspectratio,width=0.45\linewidth]{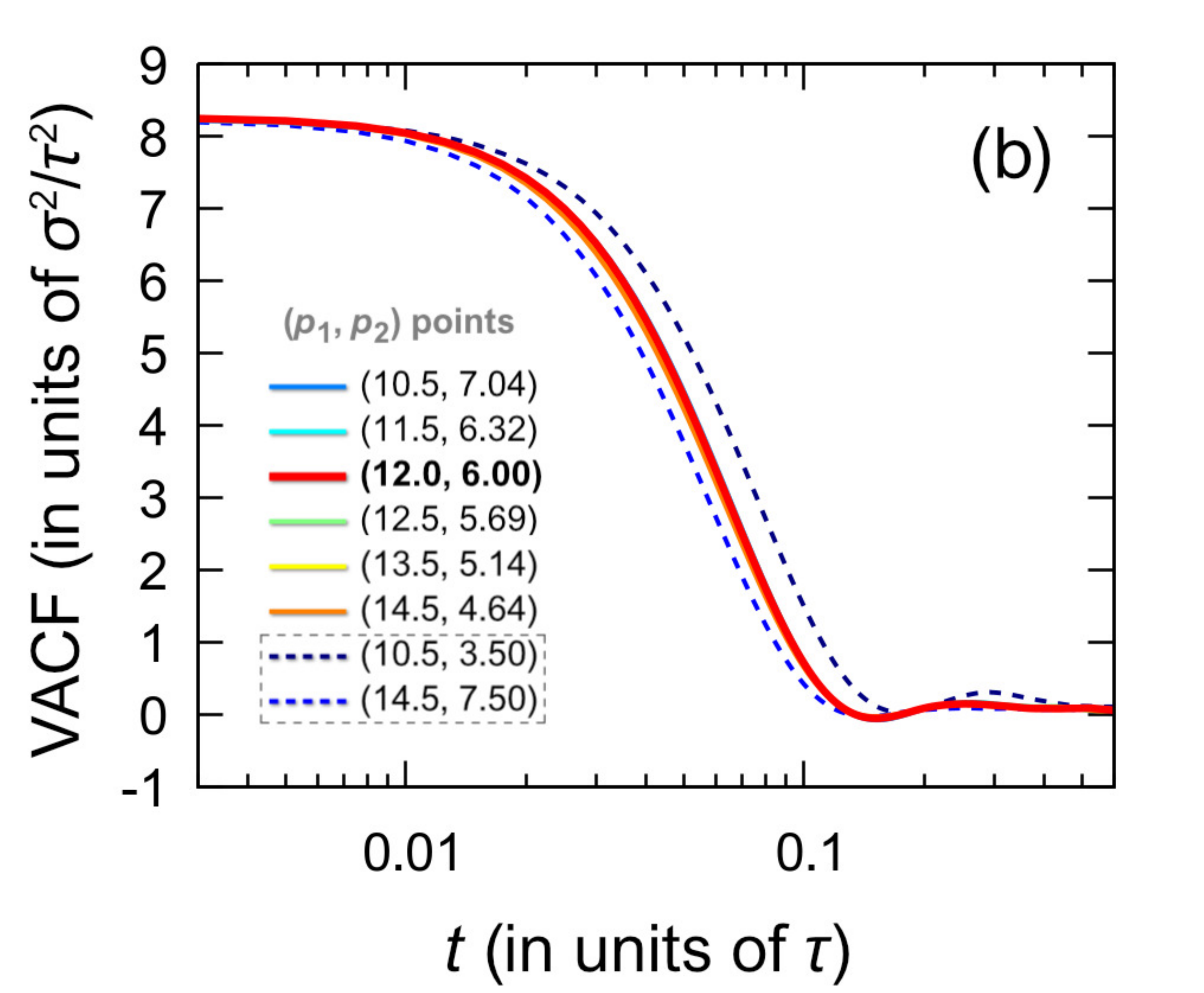}
    \includegraphics[keepaspectratio,width=0.45\linewidth]{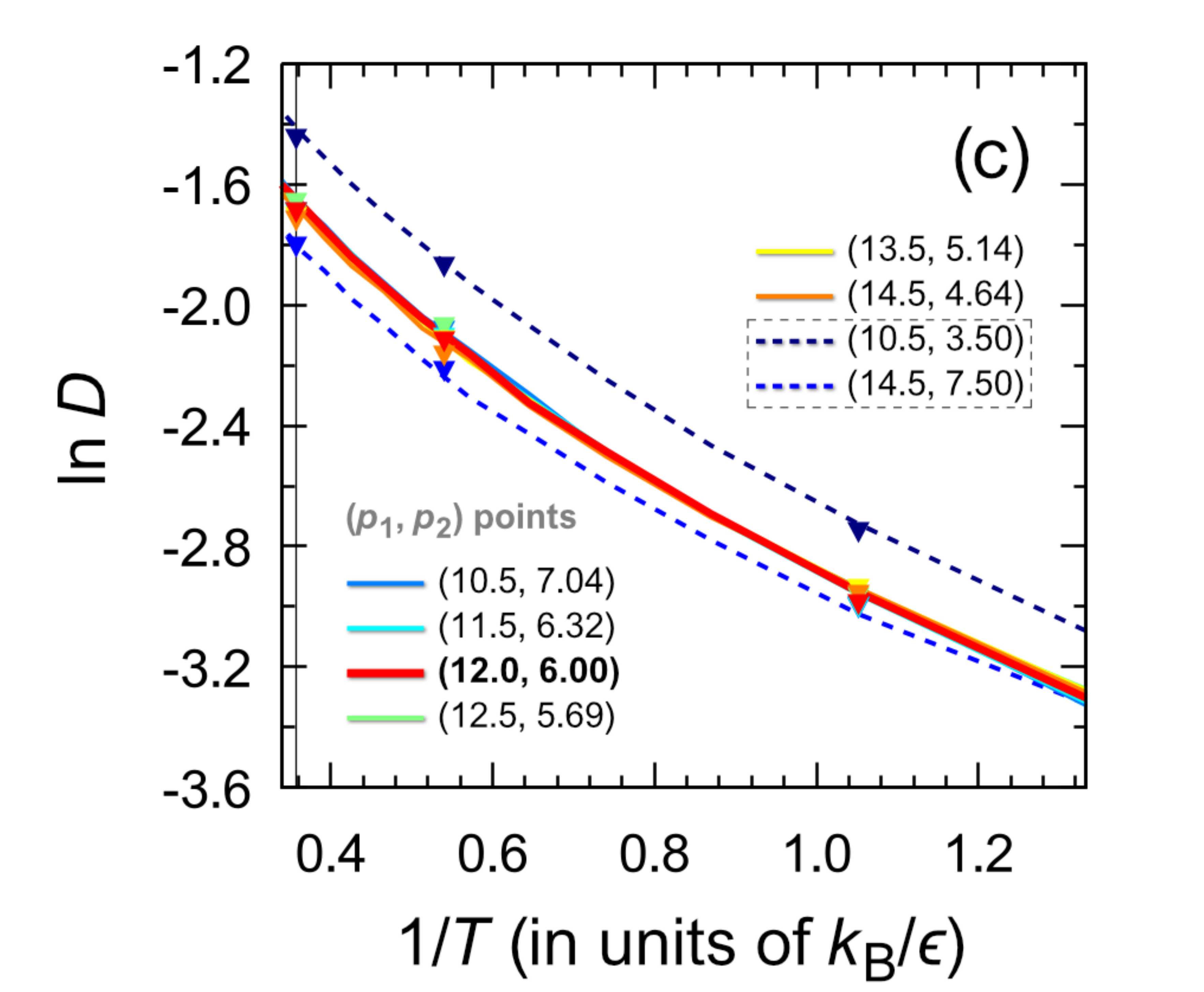}
\caption{Physical characteristics computed by means molecular dynamics simulations with the Mie potentials for
the various eight $(p_{ 1 }, p_{ 2 })$ points.
(a) Dynamic structure factor for the wavenumber $k = (0.68 \pm 0.02)\, \sigma^{ -1 }$, i.e. for $0.1\, k_{ m }$, where $k_{ m }$ is the position of the main maximum of the static structure factor $S(k)$. The time unit $\tau$ is defined as $\tau = \sqrt{m\sigma^{ 2 }/\epsilon}$.  (b) Velocity autocorrelation function. (c) Self-diffusion coefficient evaluated at different temperatures along the isochore $\rho^{ -1 } = 1.18\, \sigma^{ 3 }$ per particle. The results cover the temperature range $T \in [0.75, 2.95]\; \epsilon/k_{ B }$.  The solid lines are the results based on the mean square displacement; the markers correspond to the results obtained using the Kubo-Green relation.
}
    \label{fig: dynamics}
\end{figure}
where $W(\ldots)$ is the Lambert $W$ function~\cite{Lambert_function}. As can be seen from Fig.~\ref{fig: landscape_R}, Eq.~(\ref{eq: correspondence}) accurately reproduces the location of the valley bottom, where the function~$R(p_{ 1 }, p_{ 2 })$ takes the minimum values and the best match between the Mie and Lennard-Jones potentials is found.
Thus, at least for the range defined by $p_{ 1 } \in [10.5, 14.5]$ and $p_{ 2 } \in [4.6, 7.0]$, the Mie potential with parameters $p_{ 1 }$ and $p_{ 2 }$ satisfying Eq.~(\ref{eq: correspondence}) reproduces the interparticle interactions almost identical to the Lennard-Jones ones.

All structural features of the systems along the line containing points I-VI are practically identical. This is well confirmed by the radial distribution functions presented in Fig.~\ref{fig: structure_potential}(b). It is noteworthy that even when the repulsive and attractive contributions of the Mie potential $u(r)$ mutually exceed the Lennard-Jones ones within certain limits, this can nevertheless lead to a radial distribution function $g(r)$ typical of a Lennard-Jones fluid. This is seen, for example, for the case of point VIII with $p_{ 1 } = 14.5$ and $p_{ 2 } = 7.5$ shown in Fig.~\ref{fig: structure_potential}(b).

As known, the transport properties and collective dynamics of particles are extremely sensitive to all the nuances of interparticle interactions~\cite{Gallo2002,Fomin_Yu,Mokshin_TMP_2015,Mokshin/Galimzyanov_JCP_2014,Mokshin/Galimzyanov_JPCM_2018,Khusn_2018,Khusn_2016,Mokshin_Zabegaev_2011}. The six Mie potentials corresponding to six points I-VI produce typical Lennard-Jones-like collective particle dynamics. This is confirmed by the spectra  of the dynamic structure factor $S(k, \omega)$ calculated by means of molecular dynamics simulations with the different Mie potentials. The widths of all peaks and the positions of high-frequency peaks in the spectra of the dynamic structure factor, shown in Fig.~\ref{fig: dynamics}(a) for the case of the wavenumber $k = (0.68 \pm 0.02)\; \sigma^{ -1 }$ as an example, are practically the same. Insignificant differences are observed only in the intensities of the central (Rayleigh) and high-frequency (Brillouin) components. The coincidence of the spectra of the dynamical structure factor indicates that the collective dynamics of these systems will be also identical to the Lennard-Jones one. The results of Fig. \ref{fig: dynamics}a directly indicate that all of these six Mie potentials should reproduce the sound propagation velocity, sound attenuation, and thermal conductivity typical of a Lennard-Jones fluid.
The sound velocities for these Mie systems, estimated from molecular dynamics simulations, are as follows\footnote{The sound velocity $v_{ s }$ was evaluated from the dispersion law $\omega_{ c } = \lim_{k \to 0} v_{ s } k$, where $\omega_{ c }$ is the location of the Brillouin peak of the dynamic structure factor $S(k, \omega)$ at the generalized hydrodynamics low-$k$ regime.}:
\begin{eqnarray}
    v_{ s } = & (6.93 \pm 0.33)\, \sqrt{\epsilon/m} & \text{for the points I-VI}, \nonumber \\
    v_{ s } = & (8.11 \pm 0.2)\, \sqrt{\epsilon/m} & \text{for the point VII}, \nonumber
\end{eqnarray}
and
\begin{equation*}
    v_{ s } = (6.42 \pm 0.2)\, \sqrt{\epsilon/m}\quad \text{for the point VIII}. \nonumber
\end{equation*}
For the points I-VI, the same sound velocities, as well as the similar shapes of the high-frequency peak of the dynamic structure factor, indicate the same character of the propagation and attenuation of acoustic-like waves in the systems corresponding to the points I-VI. It is noteworthy that the velocity autocorrelation function, which characterizes the single-particle dynamics, also turns out to be the same for the potentials I-VI [see Fig.~\ref{fig: dynamics}(b)]. Considering these results on the structure and particle dynamics, it can be expected that the Mie potentials (I-VI) are able even for a range of temperatures to reproduce the physical properties typical of the Lennard-Jones system. Indeed, as can be seen in Fig.~\ref{fig: dynamics}(c), the self-diffusion coefficient $D$ calculated on the basis of the molecular dynamics simulations data along the isochore ($\rho^{ -1 } = 1.18\, \sigma^{ 3 }$ per particle) over a wide temperature range turns out to be the same for these systems.
Thus, the Mie potentials I-VI lead to the single-particle dynamics and mass transport properties identical to those for the Lennard-Jones fluid.

\section{Conclusion \label{sec: conclusion}}

The development of machine learning methods provides opportunities for solving physical problems that have been considered unsolvable for a long time. This class of problems includes the problem of restoring an interparticle interaction potential directly from empirical information about the physical properties of the system under study at the microscopic level. A necessary condition is that it should be possible to calculate analytically or numerically these physical properties on the basis of known interparticle interaction potential. For almost all (relevant) physical characteristics, this condition is satisfied. For simplest physical systems, the corresponding analytical equations are known from statistical mechanics, and these equations can only be solved in the direction ``\textit{from potential to characteristic}'', but not \textit{vice versa}. In general, physical characteristics can be calculated using molecular dynamics simulations for a given interparticle interaction potential; and this is also possible only in one direction: ``\textit{potential $\to$ physical characteristic/propery}''. Nevertheless, a very important point is what kind of information about the physical properties of the system under consideration will be optimal and sufficient to determine the potential. Would information about thermodynamic properties (e.g., equilibrium pressure, density) or about structure of the system, or about particle dynamics and transport properties of the system be suitable and sufficient?

The main result of this study is a differential evolution algorithm adapted in such a way as to restore the  interparticle interaction potential from information about the static structure of a disordered many-particle system. The input quantity of the algorithm is the radial distribution function $g(r)$, which may initially be obtained from experiments on microscopy and diffraction. The relative simplicity of calculating the function $g(r)$ by means of molecular dynamics or Monte-Carlo simulations with a taken potential makes this algorithm computationally efficient.  However, if necessary, the algorithm can be modified for the case of potential recovery from the physical properties experimentally determined for some range of temperatures and/or pressures/densities  (for example, from the self-diffusion coefficient measured by the nuclear magnetic resonance method). In this case, it would only be necessary to perform an appropriate generalization of the target function associated with a given measurable physical property.

In this study, the efficiency of the algorithm is tested on two problems. First, we considered a test optimization problem related to finding a global minimum of the Ackley function. It was found that the algorithm correctly finds this global minimum of the this function and thereby demonstrates its efficiency. Second, for the case of the Lennard-Jones system, the algorithm recovers the interparticle interaction potential solely from the known radial distribution function for some thermodynamic state. Thus, optimization by a structural characteristic turns out to be sufficient to recover the potential, and no corrections for other characteristics that take into account more subtle physical effects are required. There are no obvious restrictions on the applicability of the method to any specific systems. For monatomic or few-component systems, it can be applied directly.

The results of this study address the question of whether there is always a one-to-one correspondence between the potential, the structure and other physical properties of a system? For a system of soft particles interacting through attraction, we get the following answer. The overall effective interparticle interaction does indeed produce specific structural, dynamic, and transport properties. However, in turn, this resulting effective interparticle interaction may be a result of various combinations of attractive and repulsive contributions. This is well revealed in the example of the Lennard-Jones system considered in this study. Namely, the same structure, collective particle dynamics, and self-diffusion typical of a Lennard-Jones fluid can be generated by a whole family of interparticle interaction potentials realized under the same thermodynamic conditions. The exponents of the repulsive and attractive contributions in the potentials can take values from fairly wide intervals and these exponents turn out to be correlated.


\section*{CRediT authorship contribution statement}

\noindent \textbf{Anatolii V. Mokshin}: Conceptualization, Methodology, Analysis \& Interpretation, Writing. \textbf{Roman A. Khabibullin}: Software, Computational algorithms, Numerical simulations, Validation, Visualization.

\section*{Declaration of Competing Interest}
\noindent The authors declare that they have no known competing financial interests or personal relationships that could have appeared to influence the work reported in this paper.

\section*{Acknowledgments}
\noindent The authors are grateful to R.M. Khusnutdinoff and B.N. Galimzyanov for discussion of the results of molecular dynamics simulations and acknowledges the Foundation for the Advancement of Theoretical Physics and Mathematics ``BASIS'' for supporting the computational part of this work. This work was supported by the Russian Science Foundation (project No. $19$-$12$-$00022$).

\end{document}